\documentclass[a4paper,fleqn]{cas-dc}
\usepackage{subcaption}

\usepackage[numbers]{natbib}

\def\tsc#1{\csdef{#1}{\textsc{\lowercase{#1}}\xspace}}
\tsc{WGM}
\tsc{QE}

\begin{document}
\let\WriteBookmarks\relax
\def\floatpagepagefraction{1}
\def\textpagefraction{.001}

\shorttitle{Feasibility study to characterize the production of antineutrons in high energy $pp$ collisions, through the CEX reaction}    
\shortauthors{F. Lugo-Porras, D. M. Gomez-Coral, A. Menchaca-Rocha}  

\title [mode = title]{Feasibility study to characterize the production of antineutrons in high energy $pp$ collisions through charge exchange interactions}  



%

\author[1]{F. Lugo-Porras}[
    orcid=0009-0008-7139-3194]

\cormark[1]


\ead{lugofabiola@estudiantes.fisica.unam.mx}



\affiliation[1]{organization={Posgrado en Ciencias Físicas, Universidad Nacional Autónoma de México},
            city={CDMX},
            country={México}}

\author[2]{D. M. Gomez-Coral}[
orcid=0000-0002-9200-6607]


\ead{dgomezco@fisica.unam.mx}

\author[2]{A. Menchaca-Rocha}[
orcid=0000-0002-4856-8055]

\cormark[2]


\ead{menchaca@fisica.unam.mx}

\affiliation[2]{organization={Instituto de Física, Universidad Nacional Autónoma de México},
            addressline={Circuito de la Investigación Científica, Ciudad Universitaria}, 
            city={CDMX},
            postcode={04510}, 
            country={México}}



\cortext[1]{Corresponding author at: Posgrado en Ciencias Físicas, Universidad Nacional Autónoma de México, Ciudad de México, México.}

\cortext[2]{Corresponding author at: Instituto de Física, Universidad Nacional Autónoma de México, Circuito de la Investigación Científica, Ciudad Universitaria, Ciudad de México, 04510, México. Tel: 52 5556225060}



\begin{abstract}
Simulations to evaluate the feasibility of antineutron identification and kinematic characterization via the hadronic charge exchange (CEX) interaction $n+\bar{n}\rightarrow p+\bar{p}$ are reported. The target neutrons are those composing the silicon nuclei of which inner tracking devices present in LHC experiments (ALICE, ATLAS, and CMS) are made. Simulations of $pp$ collisions in PYTHIA were carried out at different energies to investigate $\bar{n}$ production and the expected $\bar{n}$ energy spectra. These simulations produced a decreasing power-law $\bar{n}$ energy spectra. Then, two types of GEANT4 simulations were performed, placing an $\bar{n}$ point source at the ALICE primary vertex as a working example. In the first simulation, the $E_k$ was kept at an arbitrary (1 GeV) fix value to develop an $\bar{n}$ identification and kinematics reconstruction protocol. The second GEANT4 simulation used the resulting PYTHIA at $\sqrt{s_{pp}}=13$ TeV $\bar{n}$ energy spectra. In both GEANT4 simulations, the occurrence of CEX interactions was identified by the unique outgoing $\bar{p}$. The simplified simulation allowed to estimate a 0.11\% CEX-interaction identification efficiency at $E_k = 1$ GeV. The ${p}$ CEX-partner identification is challenging because of the presence of silicon nucleus-fragmentation protons. Momentum correlations between the $\bar{n}$ and all possible $\bar{p}p$ pairs showed that $p$ CEX-partner identification and $\bar{n}$ kinematics reconstruction corresponds to minimal momentum-loss events. The use of ITS $dE/dx$ information is found to improve $\bar{n}$ identification and kinematic characterization in both GEANT4 simulations. The final protocol applied to the realistic GEANT4 simulation resulted in a $\bar{n}$ identification and kinematic reconstruction efficiency of 0.006\%, based solely on $\bar{p}p $ pair observable. Thus, the expected rate of identified and kinematically reconstructed  $\bar{n}$’s should lie in the order of 100,000 per second, illustrating the feasibility of the method. 
\end{abstract}





\maketitle

\section{Introduction}\label{intro}
\hfill

\begin{figure}
    \centering
    \includegraphics[width=7 cm, height=5cm]{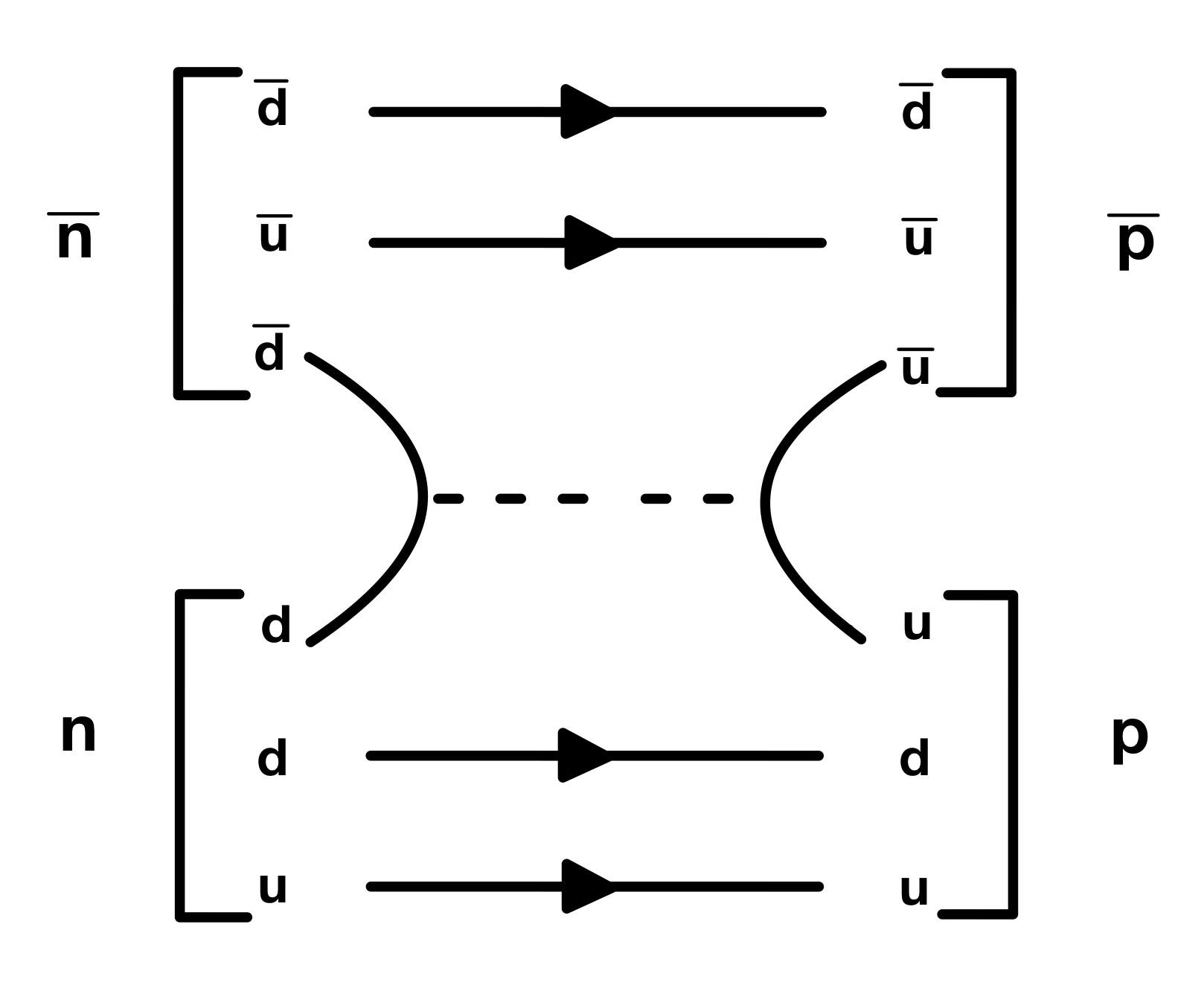}
    \caption{Charge exchange reaction (CEX) diagram for $\bar{n} + n \rightarrow \bar{p} + p$.}
    \label{cex}
\end{figure}

To a first approximation, antiproton ($\bar{p}$) and antineutron ($\bar{n}$) productions in high-energy proton-proton ($pp$) collisions should be approximately equal. Such a fact is implicit in the coalescence model \cite{coales} calculations used to characterize light anti-nuclei production in high-energy heavy-ion collisions. Yet, isotopic effects have recently been proposed \cite{ams} to significantly affect this symmetry, resulting in an enhanced $\bar{n}$ production. This is expected to have a measurable impact in the high precision cosmic ray $\bar{p}$ flux being measured by the AMS-02 experiment \cite{ams}, having as background intergalactic $pp$ collisions. This led to the inclusion of an $\bar{n}$/$\bar{p}$ asymmetry parameter, which ranges from $1.2$ to $2.0$, in the cosmic ray galactic-transport code EPOS-LHC \cite{diego}. Such an important effect should be measurable in $pp$ collisions carried out at the main LHC experiments, providing $\bar{n}$ production could be measured. Here we propose the use of the charge exchange reaction (CEX) $n+\bar{n}\rightarrow p+\bar{p}$ (see Figure \ref{cex}), which produces easy-to-identify charged particles. Indeed, this technique led to the $\bar{n}$ discovery \cite{cork}, and shortly after, to measure its mass, based on the kinematic analysis of the same reaction \cite{antin}. The feasibility of using CEX to identify and reconstruct the kinematic properties of antineutrons produced by LHC experiments via CEX interactions on their silicon inner tracking devices is explored here, taking as a working example ALICE and its Inner Tracking System (ITS).\\

 This document is organized as follows: Section \ref{simulation} describes the details of the PYTHIA and GEANT4 simulations carried out to explore the feasibility of using CEX to detect $\bar{n}$. The resulting data are presented and analyzed in section \ref{analysis}.  Momentum correlations between the $\bar{n}$ and the $\bar{p}p$ pair produced in CEX events are studied to establish viable $\bar{n}$ identification and kinematic characterization protocols, improved by using the magnitude of $dE/dx$ signals generated by CEX-induced silicon nuclei fragmentation. Conclusions are given in section \ref{conclusions}.

\section{Simulations}\label{simulation}
\hfill

 The ITS and Time Projection Chamber (TPC) detectors constitute the main ALICE charge-particle tracking devices, providing high-resolution information about interaction-vertex positions and the momentum components of the charged particle residues. The technique proposed here consists in using the ITS silicon-nuclei neutrons as CEX targets (Figure \ref{cex}) to produce easier-to-identify $\bar{p}p$ pairs. The following simulations are specifically based on the ALICE RUNs 1 and 2 ITS geometrical configuration \cite{articleALICE}, composed of 6 concentric cylindrical layers with an integrated silicon thickness of 12 mm.\\

\begin{figure}
    \centering
    \includegraphics[width=8cm, height=8cm]{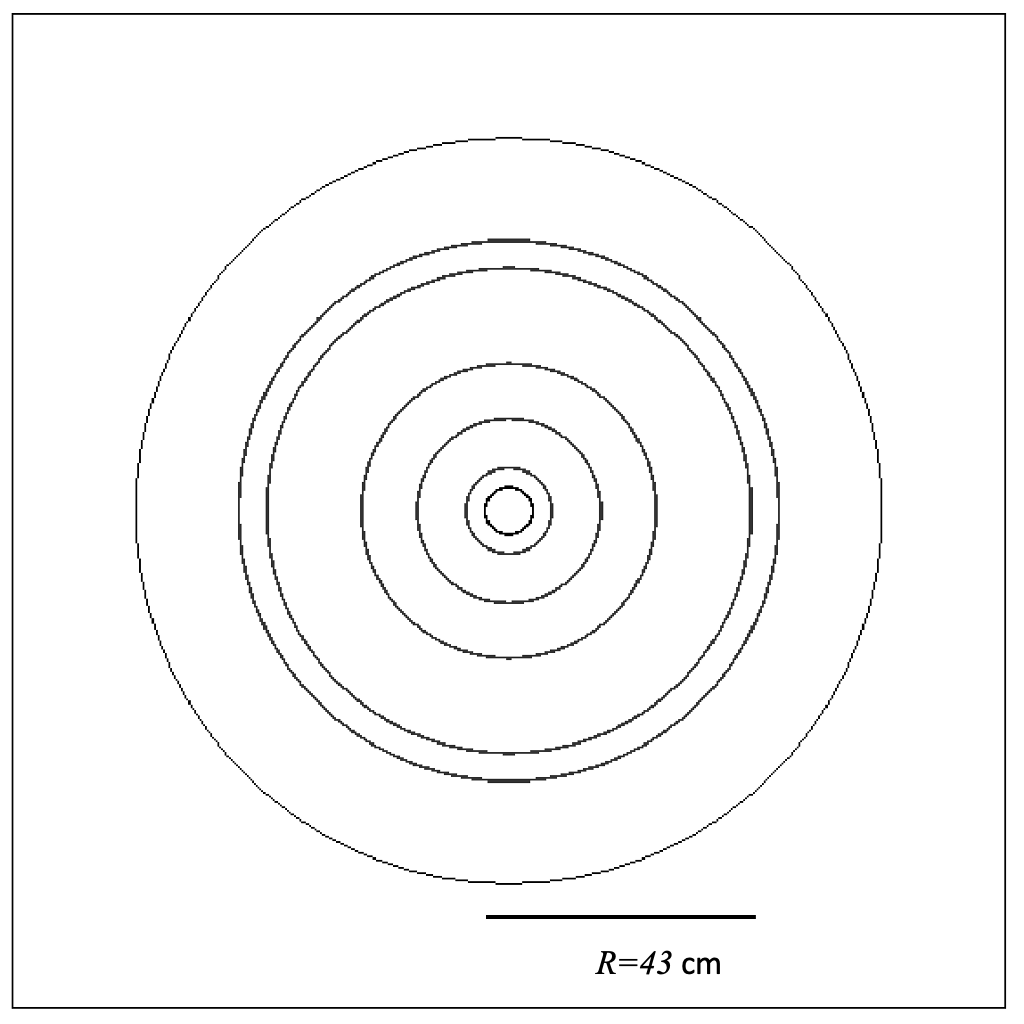}
    \caption{ITS geometry, frontal view, used in the simulations. The six inner circles represent the 2 mm \cite{spd} sensitive silicon layers. The tracker simulation limit corresponds to the seventh circle. The corresponding radii ($R$) are listed in Table \ref{tab:its}. See Figure \ref{lateral_geo} for an isometric view. }
    \label{frontal_geo}
\end{figure}

\begin{table}[h]
    \begin{center}
        \begin{tabular}{c | c | c | c  }
            \hline
            \hline
              Layer  & $R$ (cm) & $\pm Z$ (cm) & Area (m$^2$) \\
            \hline     
            1  & 3.9 & 14.1 & 0.07 \\
            2  & 7.6 & 14.1 & 0.14 \\
            3  & 15.0 & 22.1 & 0.42 \\
            4  & 23.9 & 29.7 & 0.89 \\
            5  & 38.0& 43.1 & 2.20 \\
            6  & 43.0 & 48.9 & 2.80 \\
            7  & 60.0 & 60.0 & --- \\
   
            \hline 
            \hline 
        \end{tabular}
        \caption{ALICE ITS \cite{articleALICE} layer numbers are listed from inner to outer, together with their corresponding radial ($R$) and longitudinal ($Z$) dimensions (cm), as well as the total sensitive surface area (m$^2$). The tracker simulation-limit dimensions are associated to the seventh layer.}
        \label{tab:its}
    \end{center}
\end{table}

\begin{figure}[h]
    \centering
    \includegraphics[width=8cm, height=6.7cm]{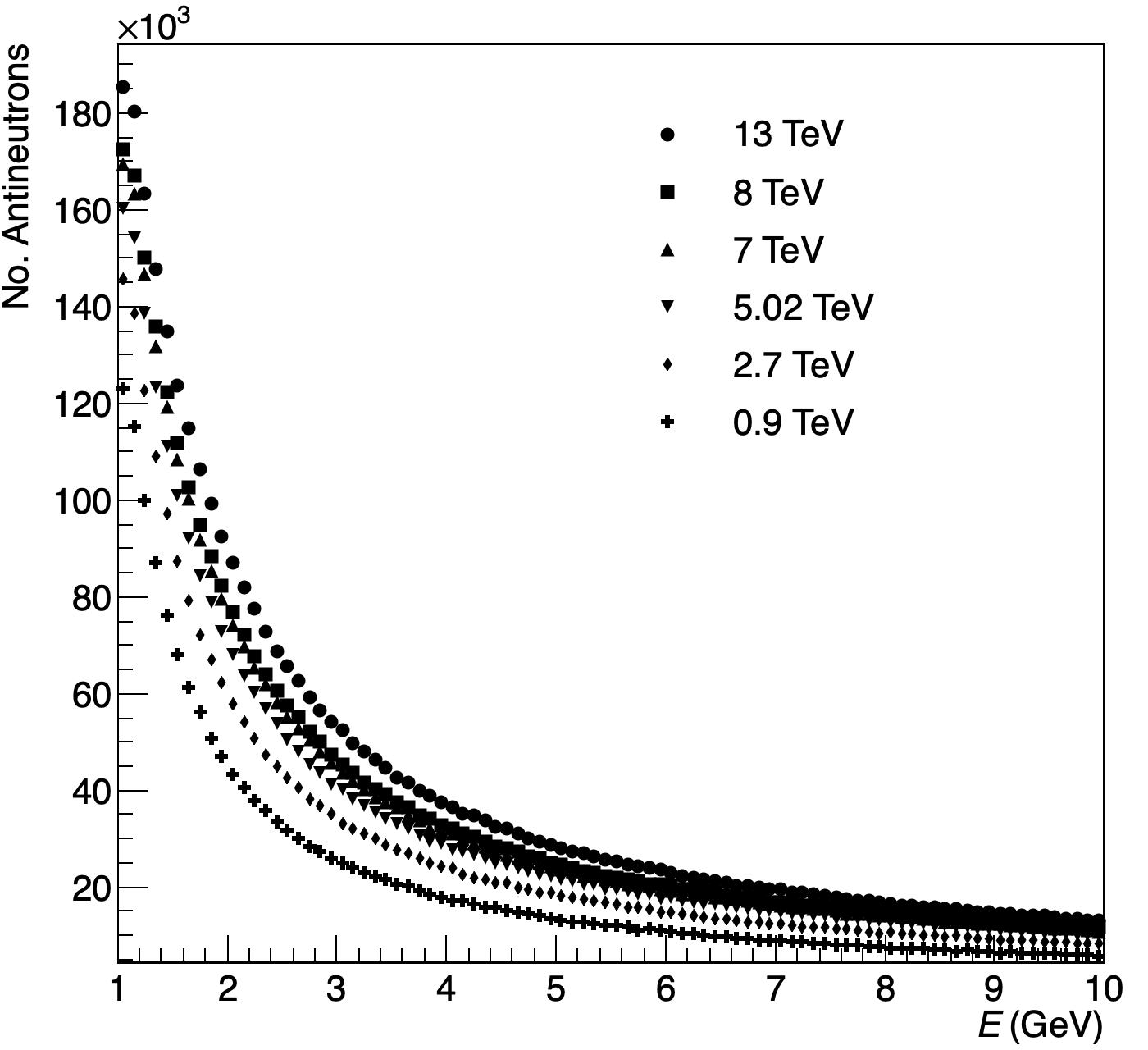}
    \caption{Total energy $E = E_k +m_0c^2$ of the antineutrons produced in events $pp$ collisions at different $\sqrt{s_{NN}}$ values generated by PYTHIA.}
    \label{energia}

\end{figure}

A first simulation was performed using the event generator PYTHIA \cite{pythia} to obtain the emitted antineutrons' energy distribution. Three million $pp$ collision events were generated at $\sqrt{s_{NN}} =$ 0.9, 2.7, 5.02, 7, 8, and 13 TeV at which ALICE \cite{pp} experimental data exists. The number of antineutrons produced per event, as well as their corresponding total energy, were recorded. The results are shown in Table \ref{tab:prod-n}, and Figure \ref{energia}, respectively. According to this simulation, at 13 TeV an average of 3 $\bar{n}$´s per collision is produced (Table \ref{tab:prod-n}). As a reference, the LHC Run 2 produced 600 million $pp$ collisions per second \cite{numerocolisiones}. Hence, on average, $10^9$ $\bar{n}$´s per second were produced at the LHC maximum energy. Figure \ref{energia} shows that the resulting $\bar{n}$ yields increase with energy, but the monotonically decreasing shape of the corresponding energy distributions remains fairly independent of the collision energy, having total energy maxima at $E\approx$ 1 GeV. Thus, in what follows, we assume that most of the $\bar{n}$ yields have kinetic energies between 0 and 2 GeV. \\

\begin{figure}
    \centering
    \includegraphics[width=8cm, height=6.5cm]{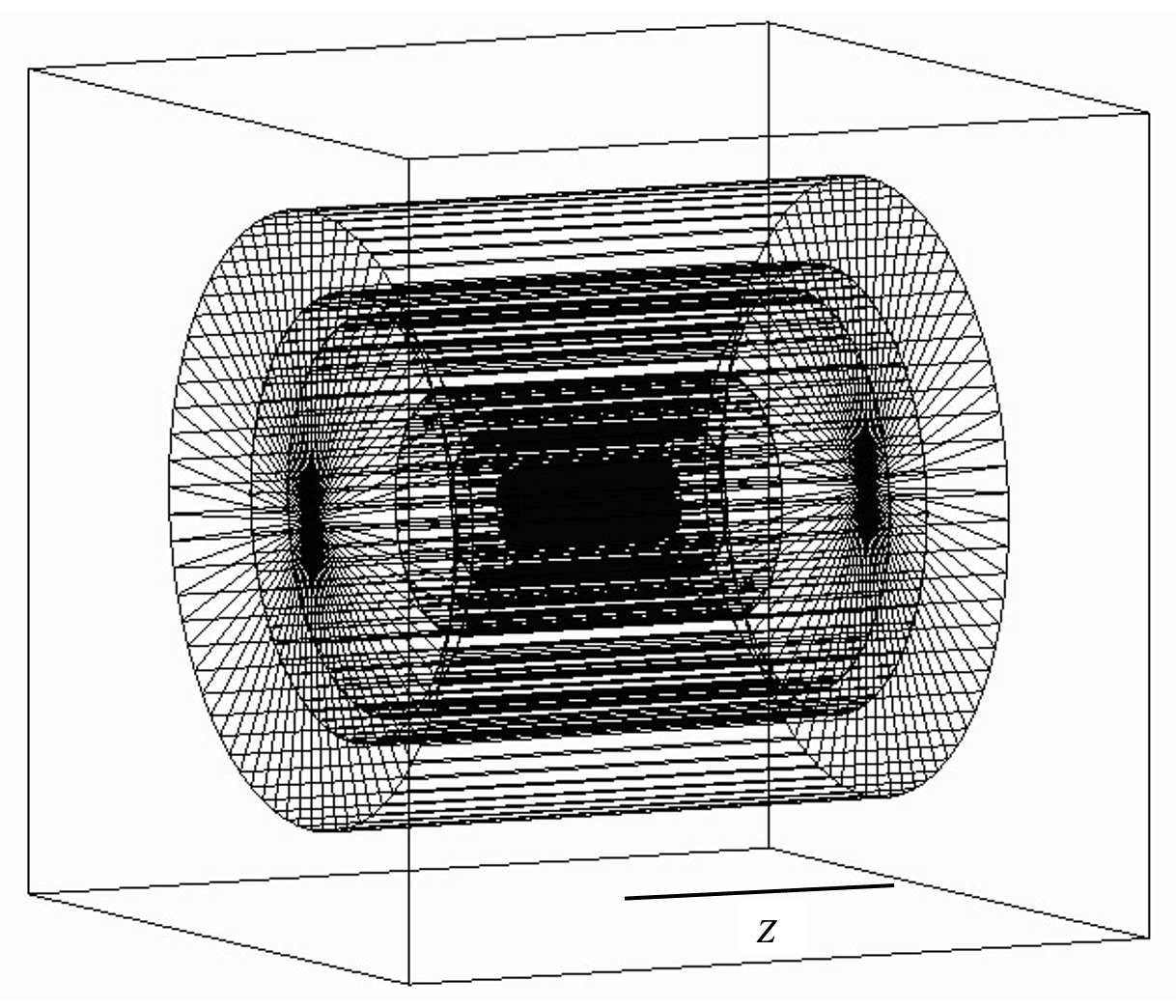}
    \caption{Isometric view of the geometry used in the GEANT4 simulations. Lines are auxiliary contours that help to improve the visualization.}
    \label{lateral_geo}
\end{figure}

\begin{figure}
    \centering
    \includegraphics[width=7cm, height=5.5cm]{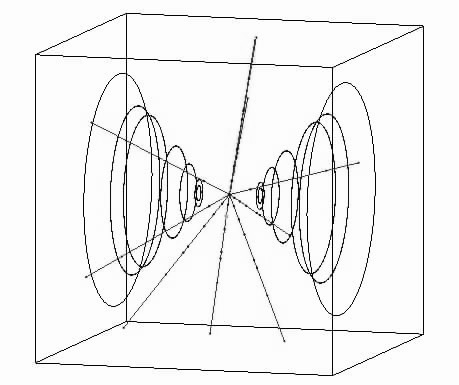}
    \caption{Isometric view of simulated $\bar{n}$ trajectories, 10 events, generated at the primary vertex. The six ITS layers $\pm Z$ borders (see Table \ref{tab:its}) are included to provide a geometrical reference.}
    \label{ejemplo_run}
\end{figure}

With the above information, a simplified GEANT4 \cite{geant4,geant4_2,geant4_3} simulation was implemented, in which a mono-energetic $E_k=1$ GeV (i.e., $c p_{\bar{n}}$ = 1.697 GeV), isotropic point $\bar{n}$ source is placed at the center of the six-layer silicon detector, hereon the primary vertex. The geometry and dimensions of the ALICE-ITS (Table \ref{tab:its}) used are shown in figures \ref{frontal_geo}, \ref{lateral_geo} and \ref{ejemplo_run}. In the GEANT4 simulations, the first ITS silicon layer is defined as the CEX target, while the remaining five layers are defined as the sensitive detectors. This, however, does not prevent CEX interactions from occurring in all six layers. 
Note that, in full (e.g., PYTHIA) $pp$ collision simulations, the presence of $\bar{p}p$ pair vertices (hereon the interaction vertices) located in outer ITS layers, with no corresponding $dE/dx$ (ionization) signals in inner ones, shall help differentiate $\bar{n}$ produced in the primary vertex from other $pp$ collision residues. This scenario, on which we are presently working, is beyond the scope of the present feasibility study, where only $\bar{n}$ are generated at the primary vertex. \\

 The present GEANT4 simulations consider an isotropic point source located at the primary vertex and emitting $\bar{n}$ of a given kinetic energy $E_k$ distribution. The simplified version considers the generation of $10^8$ fix $E_k = 1$ GeV $ \bar{n}$'s. The second  GEANT4 simulation consists of $10^7$ events, where the $\bar{n}$'s now have the actual 13 TeV PYTHIA energy distribution (see Fig. \ref{energia}). The results of both simulations are presented and analyzed in the next section. 

\begin{table}[h]
    \begin{center}
    \resizebox{8.5cm}{!} {
        \begin{tabular}{c | c | c | c | c | c | c }
            \hline
            \hline   
            $\sqrt{s_{NN}}$ TeV & 0.9 & 2.7 & 5.02 & 7 & 8 & 13 \\
            \hline 
            \hline 
            No. $\bar{n}$/collision $pp$  & 0.94 & 1.44 & 1.82 & 2.06 & 2.17 & 2.58 \\
            \hline 
            \hline 
        \end{tabular}
        }
        \caption{Average number of antineutrons produced per $pp$ collision, from PYTHIA simulations.}
        \label{tab:prod-n}
    \end{center}
\end{table}

\section{Analysis and results \label{analysis}}
\hfill

The simulated data were analyzed using ROOT \cite{root}. As a first step, $\bar{p}$´s generated by primary $\bar{n}$´s are required to generate ionization signals in at least one of the ITS layers. In the $\bar{n}$ kinetic energy domain produced by GEANT4, $n + \bar{n} \rightarrow n + \bar{p} + \pi^+$ and $n + \bar{n} \rightarrow p + \bar{p} + \pi^0$ constitute the sole background sources to the CEX interaction considered here. In order to exclude them in both GEANT4 simulations, the following $\pi$ veto was implemented. Distinguishing between the $p + \bar{p}$ and the $p + \bar{p} + \pi^0$ outgoing channels is difficult experimentally due to the short $\pi^0$ half-life ($ct = 0.26 \times 10^{-9}$ m), decaying dominantly (99\%) into two photons. However, in the $10^8$ events analyzed here, not a single one had $p + \bar{p} + \pi^0$ as an outgoing channel. Thus, the cross-section of this CEX-background channel is expected to be less than 0.1 $\mu$b, i.e., negligible compared to the above quoted $\sigma_{CEX}$ value. Concerning the $n + \bar{p} + \pi^+$ channel, 34,657 events of this type were observed in the same data sample, yielding a more significant cross-section of $5.76\pm0.12$ mb. However, in what follows, it is assumed that those events can be discriminated through the charged-particle identification of a $\bar{p} + \pi^+$ pair using downstream tracking devices (eg. the ALICE-TPC).\\

\subsection{Mono-energetic $\bar{n}$ Simulation }

Although the CEX interaction of interest is included in GEANT4 \cite{geant4_phy} as a physical process, its probability is not explicitly provided. Therefore, the corresponding $\sigma_{CEX}$ (\ref{cex}) at $E_k = 1$ GeV cross section was deduced from the $\bar{n} + ^{28}Si$ interaction simulation described next. The resulting value is, then, used to estimate the corresponding $\sigma_{CEX}$ efficiency at this energy. The simplified GEANT4 simulation, consisting of $10^8$ events, produced a total of 113,355 $\bar{p}$'s events resulting from CEX interactions. This corresponds to an identification efficiency of 0.11\%, indicating that approximately one of every 1000 antineutrons produced at the primary vertex could be identified via the $\bar{p}$ produced in the CEX reaction. This yields a CEX cross-section in silicon of $\sigma_{CEX} = 18.85\pm0.11$ mb, at this kinetic energy. Its ratio to the measured cross section for the inverse CEX reaction $\bar{p} + p \rightarrow \bar{n} + n$ on carbon (8 mb) \cite{cork}, scales approximately as their geometrical cross sections.\\

\begin{figure}[h!]
    \centering
    \includegraphics[width=8cm, height=5cm]{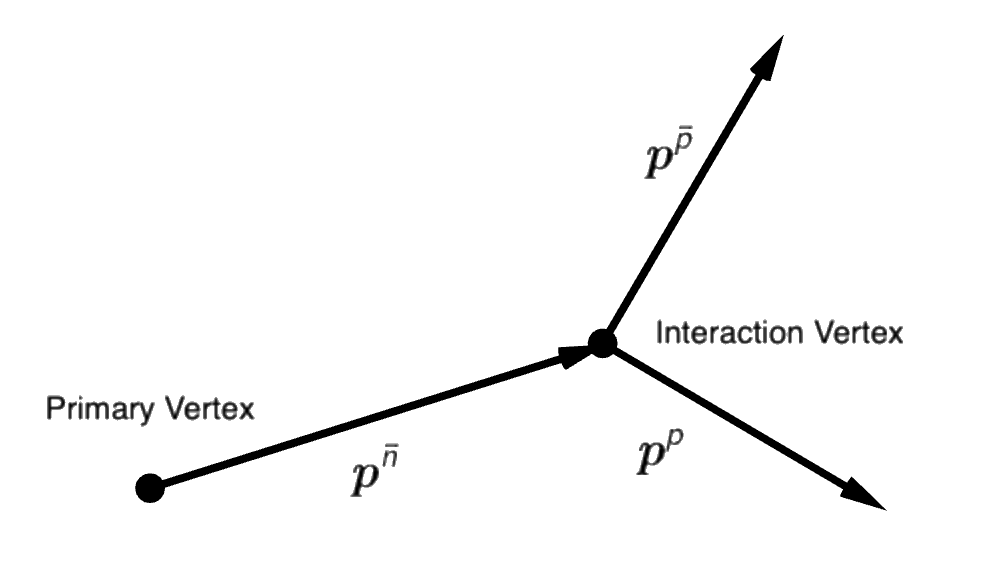}
    \caption{Momentum vectors of $p$ and $\bar{p}$ together with the interaction vertex form a single plane that, without considering energy losses due to fragmentation, should contain the primary vertex.}
    \label{geometria1}
\end{figure}

When a CEX interaction occurs, a certain amount of energy is transferred to the silicon nucleus fragments. The simulation allowed to estimate an upper limit of $dE/dx \approx 300 MeV$ ionization energy loss per event. Because of it, the $\bar{n}$ momentum may only be recovered within this limitation from a CEX-based kinematic analysis. Also, an average of 5 protons per event were found among those fragments, limiting the ability to identify the actual CEX $p$. To improve the possibility of recovering momenta information of the incident $\bar{n}$, a selection criterion was proposed to aid in identifying the CEX $p$ in each event. Based on the assumption that the primary and the interaction vertices (Figure \ref{geometria1}) can be reconstructed using downstream tracking, momentum conservation implies that the primary vertex, the CEX interaction vertex, as well as the $\textbf{p}^p$ and the $\textbf{p}^{\bar{p}}$ vectors, should lie on a common plane. In practice, energy loss to silicon fragmentation breaks this symmetry. Hence, the best one can do is to identify as the $p$ CEX-partner that fulfilling the following three conditions. First, the equation of the plane formed by the interaction vertex, the $\textbf{p}^{\bar{p}}$ vector, and the $\textbf{p}^{p}$ vector of each $p$ is determined. The most likely $p$ should be that for which the distance to the primary vertex is minimal. Second, to reduce kinematic reconstruction uncertainty, events characterized by a large loss of $\bar{n}$ energy and momentum were discarded. To do so, for every event, the quantities $E_k^{\bar{p}} + E_k^p$ and $|c(\textbf{p}^{\bar{p}} + \textbf{p}^p)|$ were calculated. The proton having the maximum values of each quantity, corresponding to the second and third conditions, was selected. Thus, the $p$ meeting all three conditions was identified as the most likely CEX $p$.\\

\begin{figure}[!tbp]
    \centering
    \includegraphics[width=8.5cm, height=6.5cm]{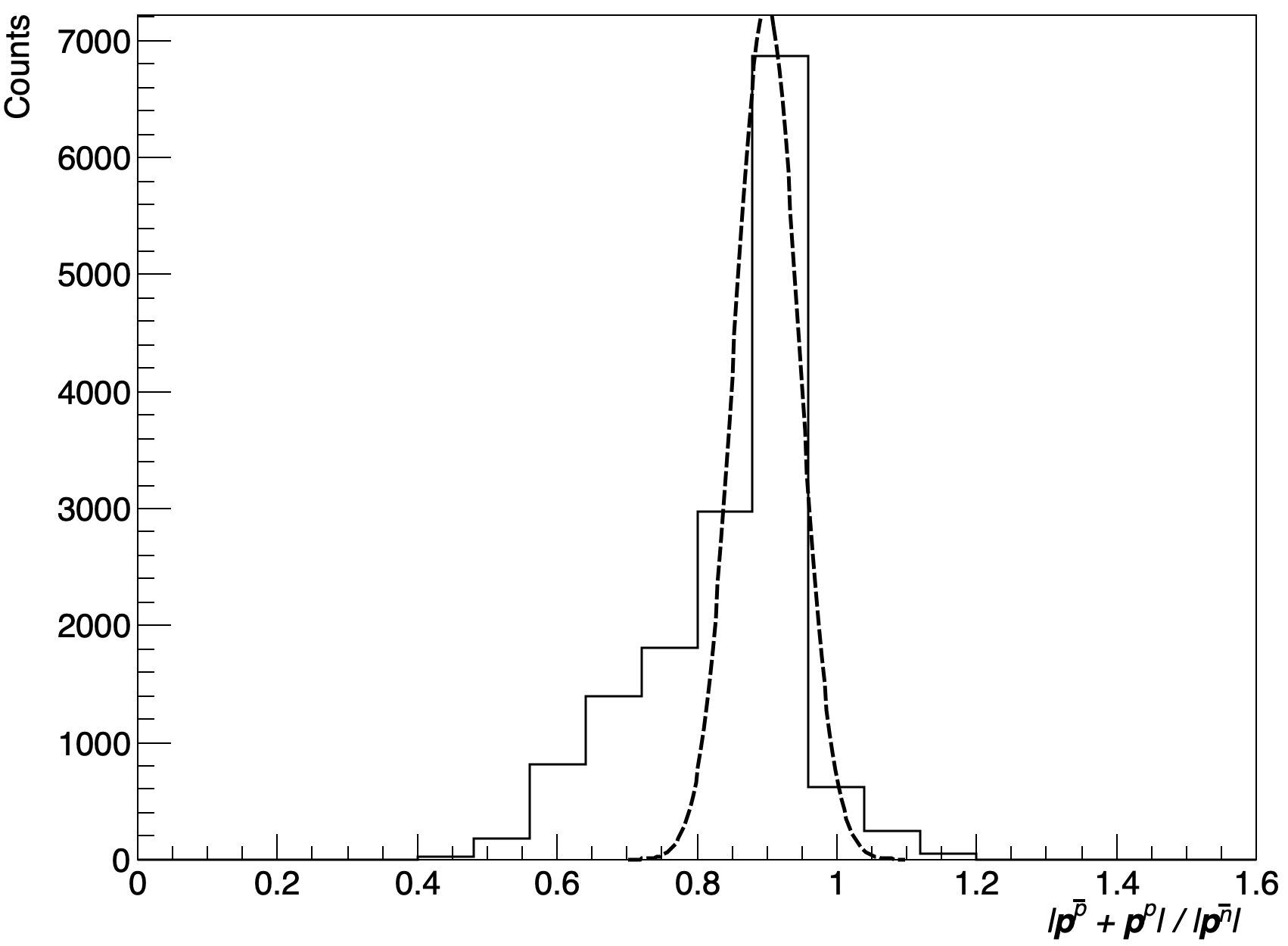}
    \caption{Mono-energetic case. The histogram shows the distribution of the reconstructed antineutron momentum magnitude $\textbf{p}^{\bar{p}} + \textbf{p}^p$ normalized by the antineutron momentum $\textbf{p}^{\bar{n}}$, coming from CEX events. The discontinuous line represents a Gaussian fit. See text. }
    \label{p_mono}   
\end{figure}

\begin{figure}[!tbp]
    \centering
    \includegraphics[width=8.5cm, height=10.5cm]{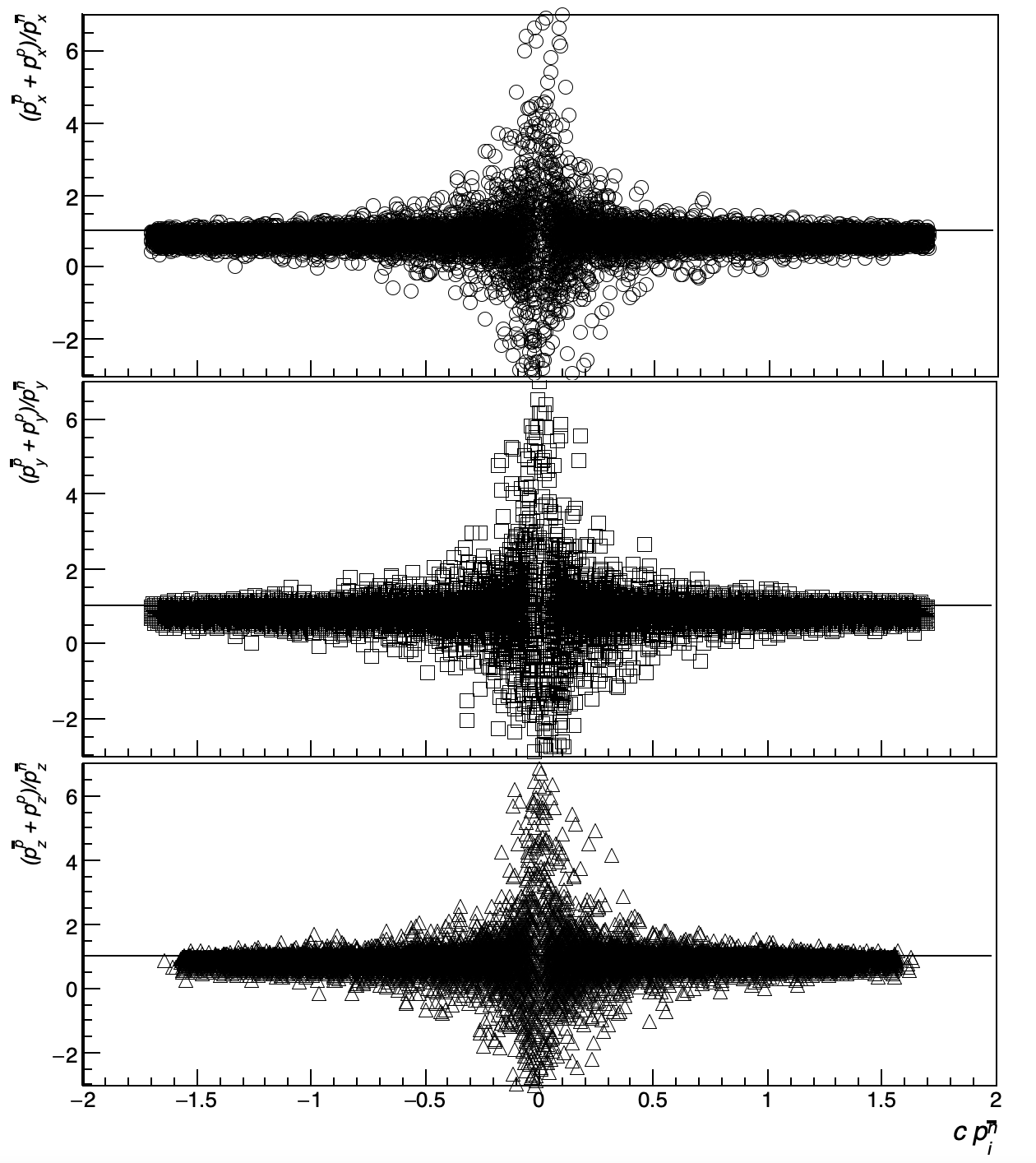}
    \caption{For events in which it is possible to identify the CEX $p$ through the method proposed, the value $p_i^{p}+p_i^{\bar{n}}$ ($i =x,y,z$) normalized by the corresponding component of the $\bar{n}$ momentum is shown versus the value $c p_i^{\bar{n}}$. Correlation factors of $\sim 0.9$ between momentum components of the CEX products and the $\bar{n}$ ones were obtained.   }
    \label{pi_results}   
\end{figure}

With the above criteria, the most likely CEX $p$ was identified in 14,969 events, i.e., 0.015\% of the mono-energetic data sample. The reconstructed $\bar{n}$ momentum norm $|\textbf{p}^{\bar{p}} + \textbf{p}^p|/ |\textbf{p}^{\bar{n}}|$ from these events is shown in Figure \ref{p_mono}, where a peak standing over a broader distribution is observed. A Gaussian fit to the peak yields a maximum at $0.90 \pm 0.05$. The underlying correlation can also be observed component-by-component in Figure \ref{pi_results}, where the normalized value $(p_i^{\bar{p}} + p_i^p)/ p_i^{\bar{n}}$ (\textit{i = x,y,z}) is plotted versus $c p_i^{\bar{n}}$. The corresponding mean values in this figure are 0.85, 0.86, and 0.85, respectively, representing the fractions of the $\bar{n}$ momentum recovered by this technique, being consistent with the $\sim90$\% obtained in Figure \ref{p_mono}. This simplified exercise already illustrates the feasibility of the technique to identify $\bar{n}$´s, allowing to count $\approx$ 0.1\% of them, while being able to reconstruct 90\% ($\pm$ 5\%) of the energy for 0.007\% of them. \\

\begin{figure}[h]
    \centering
    \includegraphics[width=8.5cm, height=6.5cm]{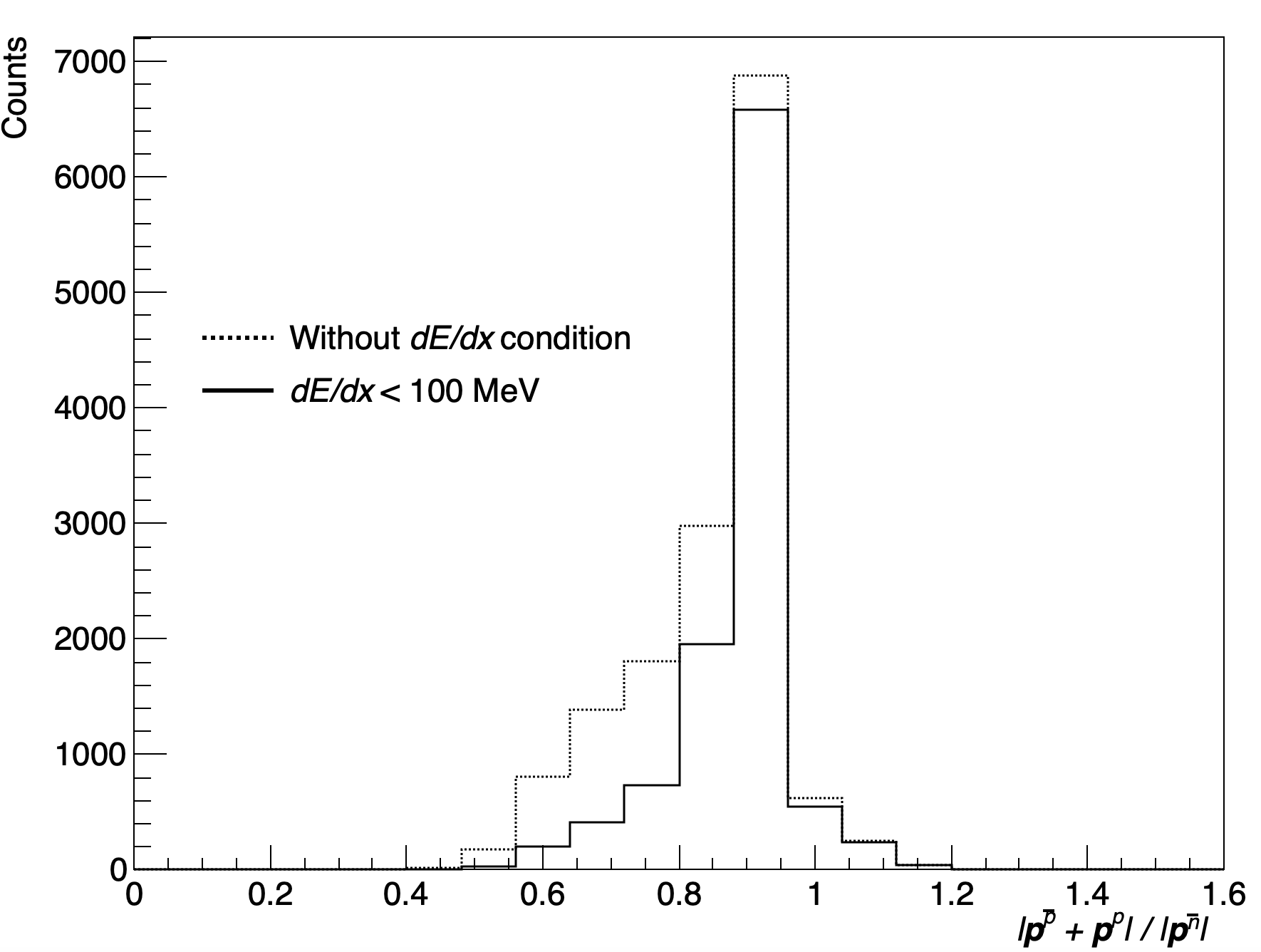}
    \caption{Mono-energetic case. Distribution of the reconstructed antineutron momentum magnitude normalized, $(\textbf{p}^{\bar{p}} + \textbf{p}^p)/\textbf{p}^{\bar{n}}$, for events in which the CEX proton can be selected and for those in which $dE/dx <100$ MeV.}
    \label{mono_dedx}
\end{figure}

The amount of energy deposited by the silicon nucleus fragments can be used to select those events characterized by a low energy transfer. Studying the corresponding $dE/dx$ signals in events where the most likely CEX $p$ identification was possible, it was empirically determined that selecting events with $dE/dx > 100$ MeV improved the kinematic reconstruction with minimal reconstructed $\bar{n}$ yield loss. These results are shown in Figure \ref{mono_dedx} for the fix $E_k$ distribution. As can be seen, when comparing the results before and after applying the $dE/dx$ condition, events with $|\textbf{p}^{\bar{p}} + \textbf{p}^p| / |\textbf{p}^{\bar{n}}|$ values between 0.5 and 0.9 were reduced, while increasing the percentage of energy that can be reconstructed from 84\% to 89\%. The percentage of events where the most likely CEX proton can be identified, which also satisfied the $dE/dx$ condition, turned out to be 0.011\%. A Gaussian fit to this distribution revealed an improved uncertainty of $\pm$ 0.04, i.e., a 20\% resolution improvement with a 26\% yield loss.

\subsection{PYTHIA 13 TeV $\bar{n}$ energy distribution simulation \label{exp_dis}}

\begin{figure}[h]
    \centering
    \includegraphics[width=8.5cm, height=6.5cm]{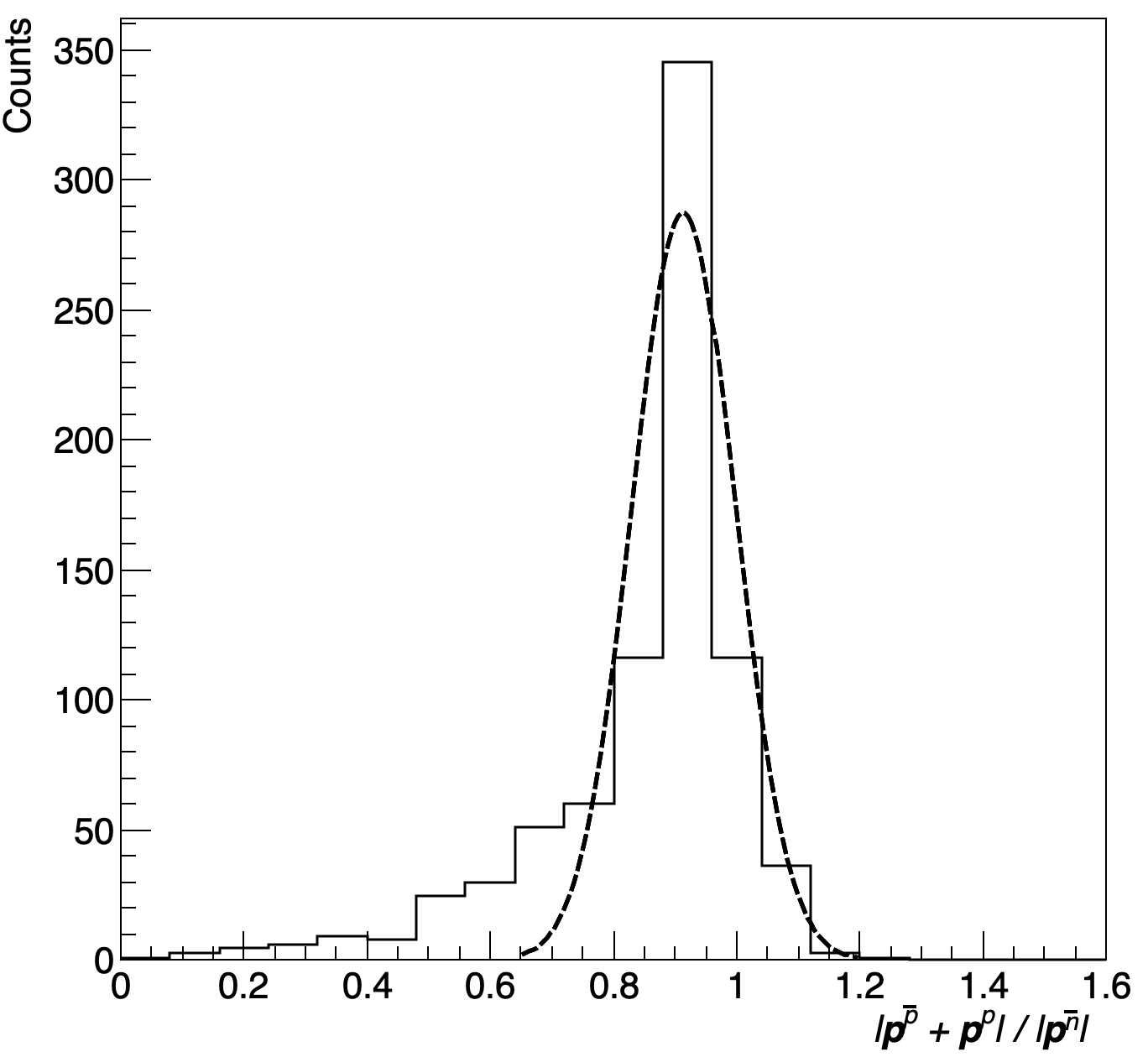}
    \caption{Realistic case. The reconstructed-to-initial antineutron momentum ratio distribution is plotted for the proton selected by the proposed method. A gaussian fit to the distribution with parameters $\mu = 0.90$ and $\sigma = 0.11$ is shown.}
    \label{p_exp}
\end{figure}

The proposed method was also applied to a realistic $\bar{n}$ source, one having the PYTHIA 13 TeV energy distribution between 0 and 10 GeV, as described in Section \ref{simulation}. This data sample comprised $10^7$ events, where 5,795 were identified as CEX interactions. From them, the most likely $p$ was identified in 815 cases. The corresponding efficiencies were 0.06\% for antineutron identification and 0.008\% for kinematic information reconstruction. The resulting normalized quantity $|\textbf{p}^{\bar{p}} + \textbf{p}^p| / |\textbf{p}^{\bar{n}}|$ is plotted in Figure \ref{p_exp}. A Gaussian fit to the peak of this distribution yields a standard deviation value of 0.08, which is assumed to be the error associated with the method proposed in this case.\\

\begin{figure}[h]
    \centering
    \includegraphics[width=8.5cm, height=6.5cm]{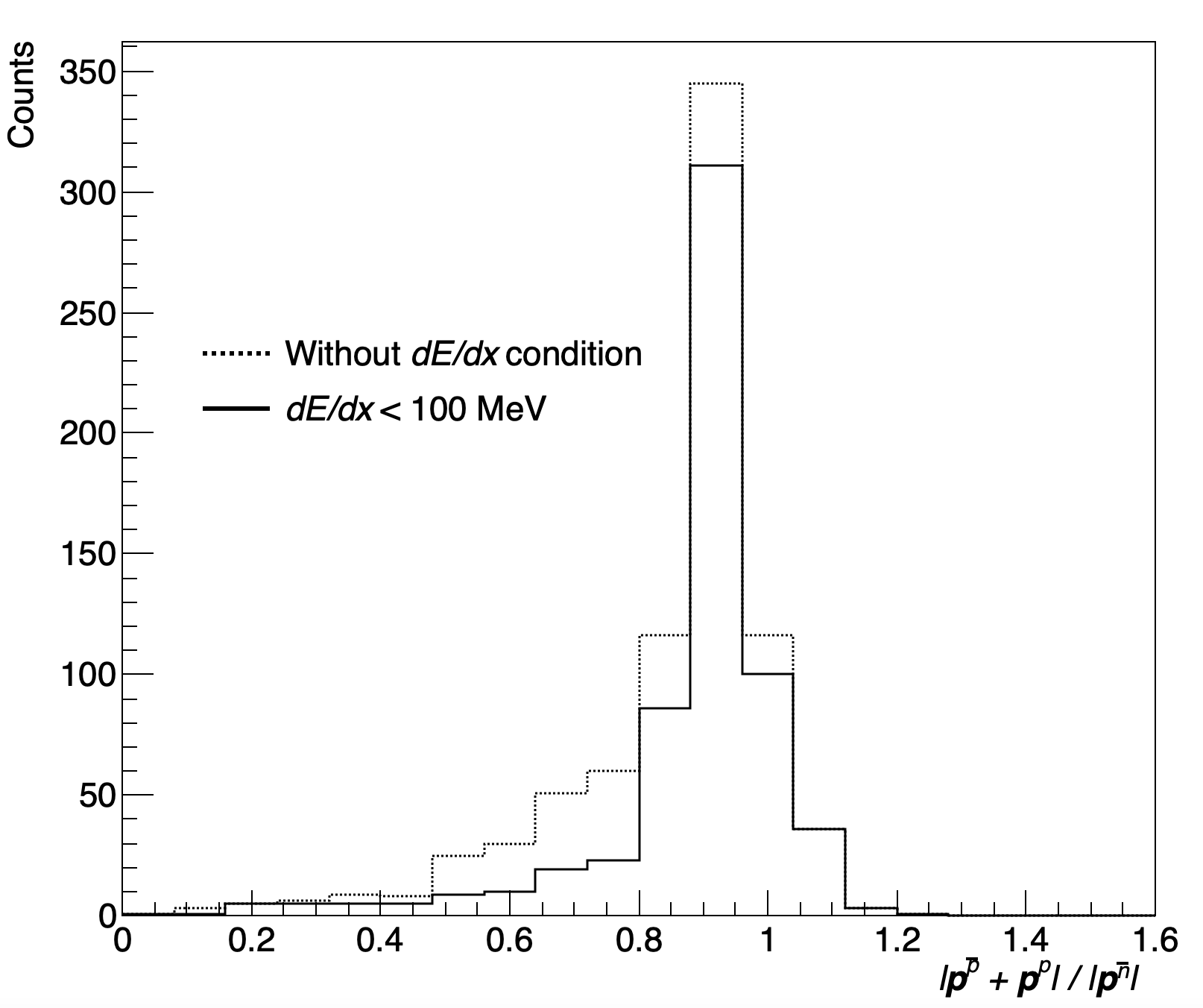}
    \caption{Realistic case. Distribution of the reconstructed antineutron momentum magnitude normalized, $(\textbf{p}^{\bar{p}} + \textbf{p}^p)/\textbf{p}^{\bar{n}}$, for events in which the CEX proton can be selected and for those in which $dE/dx <100$ MeV.}
    \label{exp_dedx}
\end{figure}

The same $dE/dx<100$ MeV condition, as in the fixed energy GEANT4 simulation, was used. The results obtained are shown in Figure \ref{exp_dedx}. Once again, comparing the results before and after applying this condition, there is a decrease in events with a  $|\textbf{p}^{\bar{p}} + \textbf{p}^p| / |\textbf{p}^{\bar{n}}|$ value between 0.4 and 0.8, increasing the percentage of energy that can be reconstructed from 85\% to 89\%. The percentage of events where the most likely CEX proton can be identified and satisfy the $dE/dx<100$ MeV condition is 0.006\%. This is similar to what was found for the mono-energetic case in figures \ref{pi_results} and \ref{p_mono}. The fit yielded an improved uncertainty of 0.06. 


\section{Conclusions \label{conclusions}}
\hfill

Simulations to estimate the identification efficiency, and kinematics-reconstruction precision, in $\bar{n}$ detection using the Charge Exchange interaction (CEX) $\bar{n}+n \rightarrow p+\bar{p}$ occurring in silicon nuclei of tracking devices in main LHC experiments, are presented. The $\bar{n}$ production rate in $pp$ collisions at the LHC was estimated via PYTHIA simulations in the $\sqrt{s_{NN}} =$ 0.9 - 13 TeV energy range, obtaining $\bar{n}/collision =$  0.9 - 2.6, respectively.  The simulated $\bar{n}$ energy spectra show maxima at the lowest energy, followed by a power law yield-decay as a function of energy. Using the ALICE-ITS detector configuration of RUN´s 1\& 2, a simplified GEANT4 simulation, locating an isotropic source emitting 1 GeV $\bar{n}$’s at the primary vertex, was carried out to estimate the CEX cross-section. Assuming all charged particles can be identified by downstream tracking, CEX-interaction identification is univocally associated to the presence of an outgoing $\bar{p}$, with no charged $\pi$'s in the event. The resulting cross section is $\sigma_{CEX} = 18.97\pm0.11$, corresponding to an $\bar{n}$ identification efficiency of 0.11\%. The competing $\bar{n}+n \rightarrow p+\bar{p} + \pi^{0}$ interaction cross section is found to be, at least, three orders of magnitude smaller.  Momentum correlations between the initial $\bar{n}$ and the outgoing $\bar{p} p$ pair were also studied. The criterion to distinguish CEX $p$'s from silicon nuclei fragmentation protons is based on momentum conservation. Yet, a high energy and momentum transfer to silicon nucleons distorts the CEX $p$ identification. The problem is reduced by discarding events with significant losses of energy and momentum, at the cost of identification efficiency.  As an example, here the $\bar{n}$ momentum could be reconstructed with an uncertainty of 5\%, while reducing the identification efficiency to 0.015\%. ITS $dE/dx$ information can also be used to reject events characterized by large momentum transfers to silicon nuclei, which produce a larger number of ionizing fragments. The addition of this selection criteria reduced the identification efficiency to 0.011\%, with an improved resolution of 4\%. A second GEANT4 simulation using a more realistic $\bar{n}$ energy distribution yielded an $\bar{n}$ identification efficiency of 0.06\% and a 0.008\% kinematic reconstruction efficiency. Should this identification and kinematic reconstruction protocol be applied to LHC Run 2 $pp$ data, the expected rate of identified and kinematically reconstructed $\bar{n}$’s should lie in the order of 100,000 per second, illustrating the feasibility of the method. 

\section*{Acknowledgement}
We would like to thank Dr. Eulogio Serradilla for the helpful discussions. This work was supported by the CONACYT [CF-2019/2042]; and the Universidad Nacional Autónoma de México [UNAM-PAPIIT IA101624].   



\bibliographystyle{elsarticle-num}

\bibliography{main}

\begin{thebibliography}{10}
\expandafter\ifx\csname url\endcsname\relax
  \def\url#1{\texttt{#1}}\fi
\expandafter\ifx\csname urlprefix\endcsname\relax\def\urlprefix{URL }\fi
\expandafter\ifx\csname href\endcsname\relax
  \def\href#1#2{#2} \def\path#1{#1}\fi

\bibitem{coales}
M.~{B{\"u}scher}, A.~A. {Sibirtsev}, K.~{Sistemich}, {Fast deuteron production
  in proton-nucleus interactions}, Zeitschrift fur Physik A Hadrons and Nuclei
  350~(2) (1994) 161--166.
\newblock \href {https://doi.org/10.1007/BF01290683}
  {\path{doi:10.1007/BF01290683}}.

\bibitem{ams}
R.~Kappl, M.~W. Winkler,
  \href{https://dx.doi.org/10.1088/1475-7516/2014/09/051}{The cosmic ray
  antiproton background for ams-02}, Journal of Cosmology and Astroparticle
  Physics 2014~(09) (2014) 051.
\newblock \href {https://doi.org/10.1088/1475-7516/2014/09/051}
  {\path{doi:10.1088/1475-7516/2014/09/051}}.
\newline\urlprefix\url{https://dx.doi.org/10.1088/1475-7516/2014/09/051}

\bibitem{diego}
D.~Gomez, Deuteron and antideuteron production in galactic cosmic-rays,
  Master's thesis, Univesidad Nacional Autónoma de México, Posgrado en
  Ciencias (Física), Instituto de Física (2019).

\bibitem{cork}
B.~Cork, G.~R. Lambertson, O.~Piccioni, W.~A. Wenzel,
  \href{https://link.aps.org/doi/10.1103/PhysRev.104.1193}{Antineutrons
  produced from antiprotons in charge-exchange collisions}, Phys. Rev. 104
  (1956) 1193--1197.
\newblock \href {https://doi.org/10.1103/PhysRev.104.1193}
  {\path{doi:10.1103/PhysRev.104.1193}}.
\newline\urlprefix\url{https://link.aps.org/doi/10.1103/PhysRev.104.1193}

\bibitem{antin}
T.~Bressani, A.~Filippi, Antineutron physics, Physics Reports-review Section of
  Physics Letters - PHYS REP-REV SECT PHYS LETT 383 (2003) 213--297.
\newblock \href {https://doi.org/10.1016/S0370-1573(03)00233-3}
  {\path{doi:10.1016/S0370-1573(03)00233-3}}.

\bibitem{articleALICE}
K.~Aamodt, A.~A. Quintana, R.~Achenbach, S.~Acounis, D.~Adamová, C.~Adler,
  M.~Aggarwal, F.~Agnese, G.~A. Rinella, Z.~Ahammed, The-ALICE-Collaboration,
  \href{https://dx.doi.org/10.1088/1748-0221/3/08/S08002}{The alice experiment
  at the cern lhc}, Journal of Instrumentation 3~(08) (2008) S08002.
\newblock \href {https://doi.org/10.1088/1748-0221/3/08/S08002}
  {\path{doi:10.1088/1748-0221/3/08/S08002}}.
\newline\urlprefix\url{https://dx.doi.org/10.1088/1748-0221/3/08/S08002}

\bibitem{spd}
A.~Kluge, G.~Aglieri-Rinella, F.~Antinori, M.~Burns, I.~A. Cali, M.~Campbell,
  M.~Caselle, S.~Ceresa, R.~Dima, D.~Elias, D.~Fabris, M.~Krivda, F.~Librizzi,
  V.~Manzari, M.~Morel, S.~Moretto, F.~Osmic, G.~S. Pappalardo, A.~Pepato,
  A.~Pulvirenti, P.~Riedler, F.~Riggi, R.~Santoro, G.~Stefanini, C.~Torcato
  De~Matos, R.~Turrisi, H.~Tydesjo, G.~Viesti,
  \href{https://cds.cern.ch/record/1089255}{{The ALICE Silicon Pixel Detector
  System (SPD)}} (2007).
\newblock \href {https://doi.org/10.5170/CERN-2007-007.143}
  {\path{doi:10.5170/CERN-2007-007.143}}.
\newline\urlprefix\url{https://cds.cern.ch/record/1089255}

\bibitem{pythia}
C.~Bierlich, S.~Chakraborty, N.~Desai, L.~Gellersen, I.~Helenius, P.~Ilten,
  L.~Lönnblad, S.~Mrenna, S.~Prestel, C.~T. Preuss, T.~Sjöstrand, P.~Skands,
  M.~Utheim, R.~Verheyen, \href{https://www.osti.gov/biblio/1873148}{A
  comprehensive guide to the physics and usage of pythia 8.3}, TBD (3 2022).
\newline\urlprefix\url{https://www.osti.gov/biblio/1873148}

\bibitem{pp}
R.~Schicker, Overview of alice results in pp, pa and aa collisions, EPJ Web of
  Conferences 138 (12 2016).
\newblock \href {https://doi.org/10.1051/epjconf/201713801021}
  {\path{doi:10.1051/epjconf/201713801021}}.

\bibitem{numerocolisiones}
Lhc $p$ collisions, taking a closer look at lhc,
  \url{https://www.lhc-closer.es/taking_a_closer_look_at_lhc/0.lhc_p_collisions},
  accessed: 2023-07-05.

\bibitem{geant4}
S.~Agostinelli, et~al.,
  \href{https://www.sciencedirect.com/science/article/pii/S0168900203013688}{Geant4—a
  simulation toolkit}, Nuclear Instruments and Methods in Physics Research
  Section A: Accelerators, Spectrometers, Detectors and Associated Equipment
  506~(3) (2003) 250--303.
\newblock \href {https://doi.org/https://doi.org/10.1016/S0168-9002(03)01368-8}
  {\path{doi:https://doi.org/10.1016/S0168-9002(03)01368-8}}.
\newline\urlprefix\url{https://www.sciencedirect.com/science/article/pii/S0168900203013688}

\bibitem{geant4_2}
J.~Allison, et~al., Geant4 developments and applications, IEEE Transactions on
  Nuclear Science 53~(1) (2006) 270--278.
\newblock \href {https://doi.org/10.1109/TNS.2006.869826}
  {\path{doi:10.1109/TNS.2006.869826}}.

\bibitem{geant4_3}
J.~Allison, et~al.,
  \href{https://www.sciencedirect.com/science/article/pii/S0168900216306957}{Recent
  developments in geant4}, Nuclear Instruments and Methods in Physics Research
  Section A: Accelerators, Spectrometers, Detectors and Associated Equipment
  835 (2016) 186--225.
\newblock \href {https://doi.org/https://doi.org/10.1016/j.nima.2016.06.125}
  {\path{doi:https://doi.org/10.1016/j.nima.2016.06.125}}.
\newline\urlprefix\url{https://www.sciencedirect.com/science/article/pii/S0168900216306957}

\bibitem{root}
R.~Brun, F.~Rademakers, Root - an object oriented data analysis framework, in:
  AIHENP'96 Workshop, Lausane, Vol. 389, 1996, pp. 81--86.

\bibitem{geant4_phy}
Geant4, physics reference manual,
  \url{https://geant4-userdoc.web.cern.ch/UsersGuides/PhysicsReferenceManual/html/index.html},
  accessed: 2023-07-05.

\end{thebibliography}



\end{document}